\documentclass[aps,prd,floatfix,superscriptaddress,preprintnumbers]{revtex4}
\usepackage{amssymb}
\usepackage{amsmath}
\usepackage{amsfonts}
\usepackage{epsfig}             
\usepackage{graphicx}
\usepackage{tabularx}
\usepackage{bm}
\usepackage{color}
\newcommand{\seq}{\begin{subequations}}
\newcommand{\sen}{\end{subequations}}
\newcommand{\eq}{\begin{eqnarray}}
\newcommand{\en}{\end{eqnarray}}
\newcommand{\la}{\langle}
\newcommand{\ra}{\rangle}

\begin{document}

\title{Meson masses and decay constants in holographic QCD \\
consistent with ChPT and HQET} 

\author{Valery E. Lyubovitskij}
\affiliation{Institut f\"ur Theoretische Physik, Universit\"at T\"ubingen, 
Kepler Center for Astro \\ and Particle Physics, 
Auf der Morgenstelle 14, D-72076 T\"ubingen, Germany}
\affiliation{Departamento de F\'\i sica y Centro Cient\'\i fico
Tecnol\'ogico de Valpara\'\i so-CCTVal, Universidad T\'ecnica
Federico Santa Mar\'\i a, Casilla 110-V, Valpara\'\i so, Chile}
\affiliation{Millennium Institute for Subatomic Physics at
the High-Energy Frontier (SAPHIR) of ANID, \\
Fern\'andez Concha 700, Santiago, Chile}
\author{Ivan Schmidt}
\affiliation{Departamento de F\'\i sica y Centro Cient\'\i fico
Tecnol\'ogico de Valpara\'\i so-CCTVal, Universidad T\'ecnica
Federico Santa Mar\'\i a, Casilla 110-V, Valpara\'\i so, Chile}

\date{\today}

\begin{abstract}

We focus on the chiral and heavy quark mass expansion 
of meson masses and decay constants. We propose a light-front  
QCD formalism for the evaluation of these quantities, consistent 
with chiral perturbation theory and heavy quark effective theory. 

\end{abstract}

\maketitle

\section{Introduction} 

In Refs.~\cite{Gutsche:2012ez,Gutsche:2017oro} we studied the chiral properties and 
heavy quark mass behavior of masses and decay constants of mesons and tetraquarks. 
We proposed longitudinal light-front wave functions (LFWFs) of mesons, which helps 
to provide a systematic chiral expansion of masses and decay constants of light 
pseudoscalar mesons $P=\pi, K, \eta$ and heavy hadrons/tetraquarks 
containing $u, d, s$ quarks. In particular, the longitudinal part of the LFWF was 
constructed in terms of current quark masses, which  helped to introduce the 
mechanism of explicit chiral symmetry breaking.
The idea of the current quark mass dependence of the longitudinal
LFWF was originally proposed in two-dimensional
large $N_c$ QCD~\cite{'tHooft:1974hx}, later was used
in the context of the two-dimensional massive Schwinger
model~\cite{Bergknoff:1976xr,Ma:1987wi,Mo:1992sv}, and 
then was applied in holographic QCD~\cite{Gutsche:2012ez,Gutsche:2017oro,%
Chabysheva:2012fe,Forshaw:2012im}.
In addition, in Refs.~\cite{Gutsche:2012ez,Gutsche:2017oro}, for the case 
of LFWFs of heavy hadrons and tetraquarks, we implemented 
the constraints of heavy quark effective theory (HQET) and potential models 
for heavy quarkonia. In particular, in the heavy quark limit $m_Q \to \infty$, we 
reproduced the mass splitting and scaling of leptonic decay constants of 
heavy-light mesons and heavy quarkonia. 

Recently, in Refs.~\cite{Li:2021jqb}-\cite{Li:2022izo}, 
our ideas were further developed by the
construction of the so-called longitudinal potential, which produces longitudinal wave functions
of hadrons. We feel that we can improve the construction of such potentials, by requiring a more  
exact correspondence with chiral perturbation theory (ChPT)~\cite{Weinberg:1978kz,Gasser:1983yg},  
which is the low-energy limit of quantum chromodynamics (QCD). In particular, such correspondence 
requires that the chiral Lagrangian/Hamiltonian must vanish in the limit of vanishing current 
quark masses of light $u, d, s$ quarks. Notice that this is not a case in the formalisms proposed 
in Refs.~\cite{Li:2021jqb}-\cite{Li:2022izo}. There are a few conditions that should be imposed. 
First of all, the quark condensate $B$ is a Lorentz invariant quantity [the vacuum expectation of 
scalar quark operator $B = |\langle 0|\bar q q|0\rangle|/(2 F_\pi^2$), where $F_\pi$ is the pion 
leptonic decay constant] with no preference of transverse or longitudinal direction, i.e. 
it obeys rotational invariance. Second, the quark condensate in QCD and ChPT 
is defined as the partial derivative of the generating functional (or Lagrangian/Hamiltonian)
with respect to current quark mass. This means that the quark condensate must be included into the
holographic Hamiltonian in such a way that the derivative of the Hamiltonian with respect to the
current quark mass gives the condensate. The solution is clear.
One should add the chiral mass term $H_\chi = {\cal M} \, B$ into the holographic Hamiltonian,  
where ${\cal M} = {\rm diag}\{m_u,m_d,m_s\}$ is the mass matrix of light $(u,d,s)$ quarks,  
which are the constituents of the respective pseudoscalar meson $P=\pi,K,\eta$. 
Such modification of the holographic Hamiltonian guarantees that its partial derivative 
with respect to the current quark masses leads to the condensate. Another point, which requires one 
to reconsider the formalisms developed in Refs.~\cite{Li:2021jqb}-\cite{Li:2022izo}, 
is the condition that the dependence on quark condensate in a Lagrangian/Hamiltonian should vanish 
in the chiral limit (i.e., when current quark masses vanish, $m_i \to 0$). 

The main objective of the present paper is to extend our ideas proposed and developed in 
Refs.~\cite{Gutsche:2012ez,Gutsche:2017oro} and derive the Hamiltonians and  
equations of motion (EOMs) producing masses and leptonic decay constants 
of light mesons and mesons containing heavy quarks, consistent with ChPT and HQET. 

The paper is organized as follows.
In Sec.~II, we present our formalism. We consistently study the chiral expansion 
of light meson masses and leptonic decay constants. 
Then, we extend our analysis on mesons containing heavy 
$c$ or $b$ quarks and derive a heavy quark mass expansion of their masses and 
lepton decay constants. Finally, Sec.~III contains our conclusions.

\section{Framework}

As we stressed in the Introduction, the task of deriving the longitudinal Hamiltonian (potential)  
in the context of light-front QCD should necessary take into account symmetry breaking 
term. In particular, it is not correct that these symmetry breaking terms 
be generated by the longitudinal part of Hamiltonian (potential) as was proposed 
in Refs.~\cite{Li:2021jqb}-\cite{Li:2022izo}. Another important point is that the full Hamiltonian 
cannot be fully universal and must be specific for each type of meson. 
We proceed step by step, starting from light pseudoscalar mesons.

\subsection{Light mesons}

First, we define the Fock states describing the light pseudoscalar mesons --- 
quark-antiquark state $|P(q_1^i\bar q_2^j)\ra = |q_1^i\ra \, |\bar q_2^j\ra$ 
with spin-pariry 
$J^P = 0^-$, where $i$ and $j$ are the $SU(3)$ flavor indices. 
The Fock states of light pseudoscalar mesons $|P\ra$, based on the 
$SU(3)$ classification, are defined in terms of $|P(q_1^i\bar q_2^j)\ra$ as 
\eq 
& &
|\pi^+\ra = |P(u_1,\bar d_2)\ra\,, \quad 
|\pi^-\ra = |P(d_1,\bar u_2)\ra\,, \quad 
|\pi^0\ra = \frac{1}{\sqrt{2}} \, 
\Big[|P(u_1,\bar u_2)\ra - |P(d_1,\bar d_2)\ra\Big] \,, \nonumber\\
& &
|K^+\ra = |P(u_1,\bar s_2)\ra\,, \quad 
|K^-\ra = |P(s_1,\bar u_2)\ra\,, \quad 
|K^0\ra = |P(d_1,\bar s_2)\ra\,, \quad 
|\bar K^0\ra = |P(s_1,\bar d_2)\ra\,, \nonumber\\ 
& &
|\eta\ra = \frac{1}{\sqrt{6}} \, 
\Big[  |P(u_1,\bar u_2)\ra 
   +   |P(d_1,\bar d_2)\ra
   - 2 |P(s_1,\bar s_2)\ra
\Big] \,. 
\en 
Next, we define the Hamiltonian of pseudoscalar mesons, which produces their masses, as 
\eq 
\hat{H}_P = \sum\limits_{k=1}^2 \hat{H}_P^{(k)} \,,
\en 
where $k$ is the index numbering 
quarks in the pseudoscalar mesons, and $\hat{H}_P$ will be specified below.  
Such Hamiltonian obeys the light-front Schr\"odinger type EOM: 
\eq 
\hat{H}_P \, |P\ra = M_P^2 \, |P\ra 
\,, 
\en 
where $M_P^2$ is the mass of pseudoscalar meson squared. 

The master formula for the mass spectrum of pseudoscalar mesons reads: 
\eq
M_P^2 = \la P| \hat{H}_P |P \ra = \int\limits_0^1 dz \, \int\limits_0^1 dx 
\ \psi_P(z,x) \, H_P(z,x) \, \psi_P(z,x) 
\,, 
\en 
where $z$ is the holographic variable, corresponding to the scale --- fifth dimension 
in the anti de-Sitter (AdS) space, $x$ is the light-cone variable, 
$\psi_P(z,x)$ is the holographic wave function 
of the pseudoscalar meson, and $H_P(z,x)$ is the representation of 
the Hamiltonian $\hat{H}_P$ in the $(z,x)$ space. Now we specify $H_P(z,x)$  
\eq
H_P(z,x) = \sum\limits_{k=1}^2 \, 
\biggl[H_{\rm kin}^{(k)}(z,x) + H_{CF}^{(k)}(z,x) + H_\chi^{(k)} 
+ H_{I}^{(k)}(z,x) \biggr]  \,. 
\en  
Here   
\eq 
H_{\rm kin}^{(1)}(z,x)  =  - \frac{d^2}{2dz^2} + \frac{m_1^2}{x} \,, \qquad 
H_{\rm kin}^{(2)}(z,x)  =  - \frac{d^2}{2dz^2} + \frac{m_2^2}{1-x} 
\en 
are the kinetic parts of the Hamiltonian acting on 
quark $q_1^i$ and antiquark $\bar q_2^j$, respectively,  
$m_1$ and $m_2$ are the masses of the corresponding current quarks,  
\eq 
H_{CF}^{(1)}(z) = H_{CF}^{(2)}(z) = \frac{4L^2-1}{8z^2} 
\en 
are centrifugal parts, where $L$ is the angular orbital momentum,  
\eq 
H_\chi^{(1)} = m_1 \, B\,, \qquad H_\chi^{(2)} = m_2 \, B\,, 
\en 
and 
\eq 
H_{I}^{(k)}(z,x) = H_{I;T}^{(k)}(z) + H_{I;L}^{(k)}(x) 
\en  
is the interaction term. The latter conventionally splits 
into a transversal part~\cite{Gutsche:2011vb}                             
\eq 
H_{I;T}^{(1)}(z) = H_{I;T}^{(2)}(z) = \frac{U_0(z)}{2} \,, 
\quad U_0(z) = \kappa^4 z^2 - 2 \kappa^2 
\en 
and a longitudinal part $H_{I;L}^{(k)}(z,x)$. The latter was 
discussed in the series of papers~\cite{Li:2021jqb}-\cite{Li:2022izo}. 
In particular, it was proposed that the longitudinal interaction potential 
generates explicitly breaking of chiral symmetry. 
As we stressed before, this is not correct since it contradicts ChPT. 
On the other hand, in the constructions of Refs.~\cite{Li:2021jqb}-\cite{Li:2022izo} 
this longitudinal interaction Lagrangian is universal for all mesons, 
which again contradicts ChPT and HQET. We found that 
the Hamiltonian $H_{I;L}^{(k)}(x)$, in the case of light pseudoscalar mesons $P$,  
must have the form 
\eq 
H_{I;L}^{(1)}(x) = H_{I;L}^{(2)}(x) = 
- \frac{\kappa^2}{2} \, 
\biggl[ \partial_x \Big(x (1-x) \partial_x\Big) 
+ (\alpha_1+\alpha_2) (1 + \alpha_1 + \alpha_2) \biggr]  \,, 
\en 
where $\alpha_i = m_i/\kappa$ are the parameters specifying the longitudinal 
part of the meson wave function and $\kappa$ is the dilaton scale parameter 
in the soft-wall AdS/QCD approach~\cite{SW_ADSQCD}. 
We remind the reader that the total mesonic LFWF function is defined as a product of 
transversal $\varphi_T(z)$, longitudinal $f_L(x)$, and 
flavor $\chi_P$  parts 
(see details, in Ref.~\cite{Gutsche:2012ez}): 
\eq 
\psi_P(z,x) = \varphi_T(z) \, f_L(x) \, \chi_P \,. 
\en 
Transverse wave functions for mesons with arbitrary spin, 
angular orbital momentum, and radial quantum number 
can be found in Ref.~\cite{Gutsche:2011vb}.                             
As it was shown by t'Hooft in Ref.~\cite{'tHooft:1974hx}, and 
confirmed in Refs.~\cite{Bergknoff:1976xr,Ma:1987wi,Mo:1992sv}, 
the longitudinal function reads  
\eq 
f_L(x) = N x^{\alpha_1} (1-x)^{\alpha_2} \,, 
\en 
where $N$ is the normalization constant fixed 
from the condition 
\eq
1 = \int\limits_0^1 dx \, \Big[f_L(x)\Big]^2,
\en 
and the $\alpha_i$ parameters are proportional to current quark masses. 

The resulting masses of the mesons get contributions from 
the transverse part $M^2_T = 4 \kappa^2 [n + (J+L)/2]$~\cite{Gutsche:2011vb}, 
the longitudinal part $M^2_L$, and additional term encoding symmetry breaking. 
For example, in the case of pseudoscalar mesons one gets: (i) $M_L^2 = 0$ due to the 
fact that the contribution of the longitudinal potential is fully compensated by the 
contribution of the mass term in kinetic term; (ii) term $H_\chi^{(k)}$ produces 
the leading order chiral corrections consistent with ChPT by construction. 

After straightforward calculations we reproduce, for the masses of pseudoscalar mesons, 
both the Gell-Mann-Oakes-Renner and the Gell-Mann-Okubo relations:   
\eq
M_\pi^2 &=& M_{\pi^\pm}^2 \,=\, 
            M_{\pi^0}^2 \,=\, 
           2B \hat{m}\,, \quad \hat{m} = \frac{m_u + m_d}{2}\,, 
\nonumber\\
M_{K^+}^2 &\equiv& M_{K^{\pm}}^2 = B (m_u + m_s)\,, \nonumber\\ 
M_{K^0}^2 &\equiv& M_{K^0/\bar{K}^0}^2 = B (m_d + m_s)\,, \nonumber\\ 
M_\eta^2 &=& \frac{B}{3} \, \Big( 2 \hat{m} + 4 m_s \Big) \,,
\nonumber\\
4 M_K^2 &=& M_\pi^2 + 3 M_\eta^2  \,, 
\en 
where 
\eq 
M_K^2 = \frac{M_{K^+}^2 + M_{K^0}^2}{2} 
\en                                     
is the average kaon mass squared. 

In the case of vector mesons the chiral symmetry breaking corrections were consistently studied in 
Refs.~\cite{Jenkins:1995vb}-\cite{Dax:2018rvs}. In particular, it was shown~\cite{Dax:2018rvs} 
that in this case there appears the same term which explicitly breaks chiral 
symmetry for the pseudoscalar mesons, but here it shows up with an arbitrary coupling $a$: 
\eq 
H_\chi^{V; a (k)} = a \, H_\chi^{(k)} \,. 
\en                           
In addition, for singlet states $\omega$ and $\phi$ there is an additional term, 
which distinguishes them from members of the $\rho$ mesons triplet $(\rho^+,\rho^-,\rho^0)$ 
and the two $K^*$ doublets $(K^{*+},K^{*0})$ and $(K^{*-},\bar K^{*0})$. 
This second term is produced by the Hamiltonian construction, using $H_\chi$  
multiplied with an additional and independent coupling $b$,
\eq 
H_\chi^{V; b (k)} = b \, H_\chi^{(k)}\,,
\en  
which is projected between matrices $V^S$ of singlet states: 
\eq 
V^S = {\rm diag}\biggl\{\frac{\omega}{\sqrt{2}},\frac{\omega}{\sqrt{2}},- \phi\biggr\} \,.
\en 
The second term gives additional corrections, in the case of $\omega$ and $\phi$ 
states: 
\eq 
\delta M_\omega^2 = 2 b B \hat{m} \,, \qquad \delta M_\phi^2 = b B m_s \,,
\en 
Combining together the contributions of the two terms responsible for explicit chiral 
symmetry breaking, one gets for vector meson masses:
\eq 
M_\rho^2 &=& M_{\rho^\pm}^2 =  M_{\rho^0}^2 = 2 \, a \, B \hat{m}
= a \, M_\phi^2 \,, \nonumber\\
M_\omega^2 &=& 2 \, (a+b)  \, B \, \hat{m} = (a+b) \, M_\pi^2 \,, \nonumber\\
M_{K^{*\pm}}^2  &=&  a \, B \, (m_u + m_s) = a \, M_{K^\pm}^2 \,, \nonumber\\
M_{K^{*0}/\bar K^{*0}}^2 &=& a \, B \, (m_d + m_s) = a \, M_{K^0/\bar K^0}^2 
\,, \nonumber\\
M_\phi^2 &=& (2a+b) B m_s = \Big(a + \frac{b}{2}\Big) \, 
(M_{K^\pm}^2 + M_{K^0/\bar K^0}^2 - M_\pi^2)\,. 
\en 
    
Other important quantities of light pseudoscalar mesons are leptonic decay constants. 
In this respect, pion leptonic decay constant was calculated for the first time in soft-wall AdS/QCD 
in Refs.~\cite{Brodsky:2007hb,Vega:2009zb,Branz:2010ub} . In addition, in Ref.~\cite{Branz:2010ub} 
leptonic decay constants of other pseudoscalar and vector mesons composed of 
both light and heavy quarks were also calculated. 
Later on, in Ref.~\cite{Gutsche:2012ez}, the effects of current quark masses in leptonic decay 
constants of light and heavy-light mesons, and heavy quarkonia, have be investigated,
where full consistency with ChPT and HQET was achieved. In particular, 
the expression for the leptonic decay constant of pseudoscalar and vector mesons, 
in terms of the $\alpha_i = m_i/\kappa$ parameters, are given 
by the expression 
\eq\label{fM_general} 
f_M(\alpha_1,\alpha_2) = \kappa \, \frac{\sqrt{6}}{\pi} \, 
\frac{\Gamma(3/2+\alpha_1) \, \Gamma(3/2+\alpha_2)}{\Gamma(3+\alpha_1+\alpha_2)} 
\,  \sqrt{\frac{\Gamma(2+2\alpha_1+2\alpha_2)}{\Gamma(1+2\alpha_1) \Gamma(1+2\alpha_2)}} 
\,. 
\en       
At leading order of chiral expansion, the leptonic decay constant is given 
by~\cite{Brodsky:2007hb,Vega:2009zb,Branz:2010ub,Gutsche:2012ez}   
\eq 
f_M^{(0)} = \frac{\kappa \sqrt{6}}{8} \,. 
\en
One can see that, in agreement with ChPT~\cite{Gasser:1983yg,Bijnens:1998di},
the leading chiral symmetry breaking correction starts with the term linear in current quark mass: 
\eq 
f_M = f_M^{(0)} \, \biggl[ 1 + \frac{m_1+m_2}{\kappa} \, \zeta 
+ {\cal O}(m_1^2,m_2^2,m_1 m_2) \biggr] \,, 
\en 
where $\zeta =  \frac{3}{2} - \log 4$. 
In particular, for the physical states of light pseudoscalar and vector mesons, one gets 
the following expressions for decay constant, including leading result and first-order 
chiral symmetry breaking correction 
\eq 
f_{\pi^\pm} &=& \frac{\kappa \sqrt{6}}{8}  \biggl[ 1 + \frac{m_u + m_d}{\kappa} \zeta \biggr] \,, 
\nonumber\\
f_{K^{\pm}} &=& \frac{\kappa \sqrt{6}}{8}  \biggl[ 1 + \frac{m_u + m_s}{\kappa} \zeta \biggr]  
\en 
and 
\eq 
f_{\rho^\pm} &=& \frac{\kappa \sqrt{6}}{8}  \biggl[ 1 + \frac{m_u + m_d}{\kappa} \, \zeta \biggr] \,, 
\nonumber\\
f_{\rho^0}   &=& \frac{\kappa \sqrt{3}}{8}  \biggl[ 1 + \frac{2 m_u + m_d}{3 \kappa} \, \zeta \biggr] \,, 
\nonumber\\
f_{\omega}   &=& \frac{\kappa \sqrt{3}}{8}  \biggl[ \frac{1}{3} 
+ \frac{2 m_u - m_d}{3 \kappa} \, \zeta \biggr] \,,
\nonumber\\ 
f_{\phi}     &=& \frac{\kappa \sqrt{6}}{8}  \biggl[ \frac{1}{3} 
+ \frac{2 m_s}{3 \kappa} \, \zeta \biggr] \,, 
\nonumber\\
f_{K^{*\pm}} &=& \frac{\kappa \sqrt{6}}{8} \biggl[ 1 + \frac{m_u + m_s}{\kappa} \, \zeta \biggr] \,, 
\nonumber\\
f_{K^{*0}/\bar K^{*0}} 
&=& \frac{\kappa \sqrt{6}}{8} \biggl[ 1 + \frac{m_d + m_s}{\kappa} \, \zeta \biggr] \,.
\en 

\subsection{Heavy-light mesons}
 
Next we discuss the mass spectrum and leptonic decay constants of heavy-light mesons. 
In this case we consider an expansion in inverse powers of the heavy quark mass and 
prove that we have full correspondence with HQET. In the following we define by $q$ and $Q$ the
light and heavy quark, respectively. 

Here, the longitudinal potential is similar to the case of light mesons and it reads 
\eq\label{LPot_qQ} 
H_{I;L}^{(1); q\bar Q}(x) = H_{I;L}^{(2); q\bar Q}(x) 
= - \frac{\kappa^2}{2} \, 
\biggl[ \partial_x \Big(x (1-x) \partial_x\Big) 
+ (\alpha_q+\alpha_Q) (1 + \alpha_q + \alpha_Q) \biggr]  \,, 
\en 
where the $\alpha$ parameters are fixed as~\cite{Gutsche:2012ez} 
\eq 
\alpha_Q = \frac{1}{2}\,, \qquad \alpha_q = \frac{\bar\Lambda}{m_Q} 
\, \biggl[1 + \frac{m_q^2 + \bar\Lambda^2}{2 m_Q \bar\Lambda}\biggr] 
- \frac{1}{2} \,,
\en 
where $m_q$ and $m_Q$ are the masses of light and heavy quark, 
$\bar\Lambda$ is the leading (of order $\Lambda_{\rm QCD}$) and flavor independent  
correction to the heavy quark mass in the expansion of the mass of heavy-light meson 
$M_{q\bar Q}$ in HQET~\cite{Neubert:1993mb}: 
\eq\label{MqQ_expansion} 
M_{q\bar Q} = m_Q + \bar\Lambda + {\cal O}(1/m_Q) 
\en 
Due to our choice of the $\alpha_q$ and $\alpha_Q$ parameters, we 
exactly reproduce the expansion for the mass of heavy-light mesons, i.e. 
Eq.~(\ref{MqQ_expansion}). Note that this expansion is governed by the longitudinal 
potential~(\ref{LPot_qQ}).  

Now we turn to discussion of the results for leptonic decay constants $f_{qQ}$ of heavy-light mesons. 
Taking Eq.~(\ref{fM_general}) for the leptonic decay constant of a meson with arbitrary quarks 
and substituting $\alpha_1 = \alpha_q$ and $\alpha_2 = \alpha_Q = 1/2$ we get 
\eq 
f_{qQ} &=& \frac{\kappa \sqrt{6}}{\pi} \, \frac{\Gamma(3/2+\alpha_q)}{\Gamma(7/2+\alpha_q)} \, 
\sqrt{\frac{\Gamma(3+2\alpha_q)}{\Gamma(1+2\alpha_q)}} \nonumber\\ 
&=& 
\frac{\kappa \sqrt{6}}{\pi} 
\, \frac{\sqrt{r (1+r)}}{(1+r/2) (2+r/2)} \,,
\en 
where 
\eq 
r = 1 + 2 \alpha_q = \frac{2 \bar\Lambda}{m_Q} \, 
\biggl[1 + \frac{m_q^2 + \bar\Lambda^2}{2 m_Q \bar\Lambda}\biggr] 
\en 
is the small parameter of order ${\cal O}(1/m_Q)$ in which powers we can expand. 
We get: 
\eq 
f_{q\bar Q} = \frac{\kappa \sqrt{6 r}}{2 \pi} \, \biggl[ 
1 - \frac{r}{4} + {\cal O}(r^2) \biggr] \sim \sqrt{\frac{\bar\Lambda}{m_Q}}
\en 
One can see that at leading order of the heavy quark mass expansion, the decay constant 
$f_{q\bar Q}$ scales as $\sqrt{1/m_Q}$, in full agreement with HQET~\cite{Neubert:1993mb}. 
Another interesting result is that the chiral corrections appear at order $m_q^2$ and are suppressed 
in comparison with the linear $m_q$ correction, which could be induced by the
chiral Hamiltonian $H_\chi$ explicitly breaking chiral symmetry. 

\subsection{Heavy quarkonia}

Finally, we consider the heavy quark mass expansion of masses and decay constants of 
heavy quarkonia. We start by specifying the longitudinal potential for heavy quarkonia
\eq\label{LPot_QQ} 
H_{I;L}^{(1); Q_1\bar Q_2}(x) = H_{I;L}^{(2); Q_1\bar Q_2}(x) 
= - \frac{8 E^4}{(m_{Q_1}+m_{Q_2})^2}  \, 
\biggl[ \partial_x \Big(x (1-x) \partial_x\Big)  
+ (\alpha_{Q_1}+\alpha_{Q_2}) (1 + \alpha_{Q_1} + \alpha_{Q_2}) \biggr]  \,, 
\en 
and dilaton parameter as 
\eq 
\kappa_{Q_1\bar Q_2} = 
\kappa \, \biggl(\frac{\mu_{Q_1\bar Q_2}}{E}\biggr)^{3/4} \,,  
\en 
where 
\eq 
\mu_{Q_1\bar Q_2} = \frac{m_{Q_1} m_{Q_2}}{m_{Q_1}+m_{Q_2}}
\en 
is the reduced mass of the bound state composed of heavy quarks 
$Q_1$ and $Q_2$, and $\kappa$ is the parameter of dimension of mass, 
which is of order ${\cal O}(1)$, i.e. independent on heavy flavor. 
$E$ is the binding energy, which is defined as the leading correction to the heavy 
quark masses $m_{Q_1}$ and $m_{Q_2}$ in the heavy quark mass expansion of 
the mass of heavy quarkonia $M_{Q_1\bar Q_2}$: 
\eq 
M_{Q_1\bar Q_2} = m_{Q_1} + m_{Q_2} + E + {\cal O}\Big(1/m_{Q_1},1/m_{Q_2}\Big) \,
\en 
In order to get consistency with HQET 
we fix the $\alpha_{Q_i}$ parameters as~\cite{Gutsche:2012ez} 
\eq 
\alpha_{Q_i} = \frac{m_{Q_i}}{4E}\, \biggl[1 - \frac{E}{2m_{Q_i}}\biggr] 
\,. 
\en 
Now we look at the leptonic decay constants of heavy quarkonia. 
First, we consider the leptonic decay constant of heavy quarkonia composed 
of quark and antiquark of the same flavor $Q_1 = Q_2 = Q$. 
Using Eq.~(\ref{fM_general}) and substituting there 
$\alpha_{Q_1} = \alpha_{Q_2} = \alpha_Q$ one gets for leading term 
in the heavy quark mass expansion:  
\eq 
f_{Q\bar Q} &=& \frac{\kappa_{Q\bar Q} \sqrt{6}}{\pi^{3/4}} 
\, \frac{1}{(2\alpha)^{1/4}} \nonumber\\
&=& 
\frac{\kappa \sqrt{3}}{\pi^{3/4}} \, \sqrt{\frac{m_Q}{E}} 
\sim \sqrt{m_Q} 
\en 
in full agreement with HQET. 

Another interesting case is the leptonic decay constant of the $B_c^+(c\bar b)$ 
meson. Here, we apply the condition that the mass of charm quark is much smaller than 
the mass of the bottom quark $m_c \ll m_b$. In this limit one gets 
\eq 
f_{c\bar b} = 
\frac{2 \kappa \sqrt{6}}{\pi^{3/4}} \, \frac{m_c}{\sqrt{m_b E}} \sim 
\frac{m_c}{\sqrt{m_b}} \,.
\en  

\section{Conclusion} 

In this paper we continue our study of the consistency of light-front QCD motivated by 
soft-wall AdS/QCD, with ChPT and HQET. In particular, 
in Refs.~\cite{Gutsche:2012ez,Gutsche:2017oro} we preliminary studied chiral properties and 
heavy quark mass behavior of masses and decay constants of mesons and tetraquarks. 
We proposed longitudinal LFWFs of mesons and tetraquarks providing 
systematic and consistent chiral expansion of masses of mesons and tetraquarks. 
In Refs.~\cite{Gutsche:2012ez,Gutsche:2017oro} we did not specify the longitudinal 
potential, which should accompany the corresponding LFWFs. 
In recent papers~\cite{Li:2021jqb}-\cite{Li:2022izo} 
our ideas were further developed by derivation of the longitudinal potential, which 
produces masses of mesons and leading-order chiral corrections. As we stressed in 
the Introduction, Refs.~\cite{Li:2021jqb}-\cite{Li:2022izo} actually failed in the construction 
of this longitudinal potential. We claim that the source of the explicit 
breaking of chiral symmetry should be introduced following ChPT and it is a Lorenz invariant 
quantity and cannot be related to the longitudinal dynamics of the bound state in LF QCD. 
We showed how to construct the longitudinal potential in order to get consistency with ChPT 
and also with HQET. Analytical results for leading correction for meson masses and leptonic 
decay constants were derived. Also we demonstrated how to proceed in the case of heavy-light 
mesons and heavy quarkonia, to get correspondence with HQET for the expansion 
of masses of these states and the power scaling of their leptonic decay constants at 
infinitely large values of heavy quark masses. 

\begin{acknowledgments}

This work was funded by BMBF (Germany) ``Verbundprojekt 05P2021 (ErUM-FSP T01) -
Run 3 von ALICE am LHC: Perturbative Berechnungen von Wirkungsquerschnitten
f\"ur ALICE'' (F\"orderkennzeichen: 05P21VTCAA), by ANID PIA/APOYO AFB180002 (Chile),
by FONDECYT (Chile) under Grant No. 1191103,  
and by ANID$-$Millen\-nium Program$-$ICN2019\_044 (Chile).

\end{acknowledgments}

\end{document}